\documentclass[final]{svjour2}
\usepackage{graphicx}
\usepackage{rotating}
\usepackage{amssymb}
\usepackage{mathptmx}
\usepackage[numbers]{natbib}
\makeatletter
\journalname{Journal of Low Temperature Physics}

\bibpunct{}{}{,}{s}{}{,}

\begin{document}

\newcommand{\hdblarrow}{H\makebox[0.9ex][l]{$\downdownarrows$}-}
\title{Horn Coupled Multichroic Polarimeters for the Atacama Cosmology Telescope Polarization Experiment}

\author{\small R. Datta,$^{1}$ \and J. Hubmayr,$^{2}$ \and C. Munson,$^{1}$ \and J. Austermann,$^{3}$ \and J. Beall,$^{2}$ \and D. Becker,$^{2}$ \and H. M. Cho,$^{2}$ \and  N. Halverson,$^{3}$ \and G. Hilton,$^{2}$ \and K. Irwin,$^{2}$ \and D. Li,$^{2}$ \and J. McMahon,$^{1}$ \and L. Newburgh,$^{4}$ \and J. Nibarger,$^{5}$ \and M. Niemack,$^{2,6}$ \and B. Schmitt,$^{7}$ \and H. Smith,$^{1}$ \and S. Staggs$^{4}$ \and J. Van Lanen,$^{2}$ \and E. Wollack,$^{8}$}

\institute{\small $^1$Department of Physics, University of Michigan Ann Arbor, MI 48109\\
\email{dattar@umich.edu}\\
{$^2$National Institute of Standards and Technology, 325 Broadway, Boulder, CO 80305}\\
{$^3$ Department of Astrophysical Sciences, University of Colorado, Boulder, CO 80309}\\
{$^4$Princeton University Department of Physics, Jadwin Hall, P. O. Box 708, Princeton, NJ 08544}\\
{$^5$Boulder Micro-Fabrication Facility, National Institute of Standards and Technology, 325 Broadway, MS 817.03, Boulder, CO 80305}\\
{$^6$Cornell University Physics Department, 109 Clark Hall, Ithaca, New York 14853}\\
{$^7$University of Pennsylvania Department of Physics and Astronomy, 209 South 33rd St., Philadelphia, PA 19104}\\
{$^8$NASA Goddard Space Flight Center, 8800 Greenbelt Road, Greenbelt, MD 20771}}

\date{07.20.2013}

\maketitle

\begin{abstract}

Multichroic polarization sensitive detectors enable increased sensitivity and spectral coverage for observations of the Cosmic Microwave Background (CMB). An array optimized for dual frequency detectors can provide 1.7 times gain in sensitivity compared to a single frequency array. We present the design and measurements of horn coupled multichroic polarimeters encompassing the 90 and 150 GHz frequency bands and discuss our plans to field an array of these detectors as part of the ACTPol project.  


\end{abstract}

\section{Introduction}
Measurement of the faint polarized signals of the cosmic microwave background will provide a new window into the physics of inflation \cite{CMBPol:Bauman}, a measurement of the neutrino mass sum \cite{CMBPol:Smith}, and a rich assortment of other astrophysical and cosmological results.   Detecting these signals requires instruments with high sensitivity and excellent control over systematic errors.  Multichroic detectors with sensitivity to multiple spectral bands within a single focal plane element offer an avenue to achieve these goals.  We present the status of our work on horn coupled multichroic polarimeters and our plans to deploy an array of these detectors on the Atacama Cosmology Telescope (ACT) as part of the ACTPol project \cite{ACTPol:Niemack}.  The ACTPol instrument is optimized to measure CMB polarization, CMB lensing, and other secondary anisotropies.

In \S \ref{sec:desnperf} we present the design of our horn coupled polarimeters which operate in both the 90 and 150 GHz bands.  Measurements of these detectors are compared to simulations in \S \ref{sec:mcperf}.  In \S \ref{sec:arropt} we describe the array optimization and implementation on ACTPol.  This includes a discussion of our approach to broad-band AR coating the silicon lenses used in the ACTPol optical design.  We conclude in \S \ref{sec:conc} with a discussion of the future applicability of this technology and the schedule for deployment on ACTPol.

\section{Multichroic Pixel Design}
\label{sec:desnperf}

\begin{figure}
\begin{center}
\includegraphics[width=0.75\linewidth,keepaspectratio]{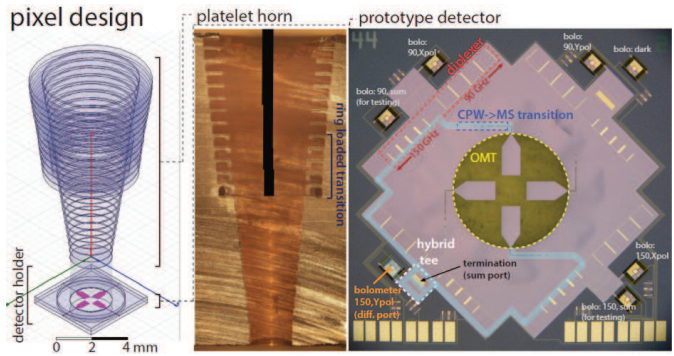}
\end{center}
\caption{{\it Left:} \small design of a single horn coupled multichroic polarimeter with labels on the major components.  {\it Center:} A photograph of a cross-section of a broad-band ring-loaded corrugated feed horn fabricated by gold plating a stack of etched silicon platelets. {\it Right:} A photograph of a prototype 90/150 multichroic detector with the major components labeled.  A description of these components is in the text.  For clarity, the path light follows to reach the bolometer corresponding to Y polarization in the 150 GHz band has been highlighted. (color figure online)}
\label{fig:pixdesign}
\end{figure}

Figure \ref{fig:pixdesign} shows a rendering of a single multichroic polarimeter with sensitivity to the 90 and 150 GHz CMB bands and photographs of a prototype horn and detector chip.  The beam pattern is defined by a broad-band ring-loaded corrugated feed horn.  The incoming light is converted into electrical signals using superconducting circuitry on a planar detector wafer that separates the incoming light according to polarization and frequency band. The signals are detected using transition edge sensor (TES) bolometers and read out using SQUIDs coupled to the TESs.

This ring-loaded\cite{Ringloadedguide} horn design is similar to a standard corrugated feed but with the addition of five ring-loaded sections that form a broad-band impedance matching transition between the round input waveguide and the corrugated output guide that defines the beam pattern.  This horn was optimized to produce a round beam with low cross-polarization over a band from 70 to 175 GHz, but has excellent performance up to 225 GHz. Prototypes (including the one that has been cut in half and photographed in Figure \ref{fig:pixdesign}) have been fabricated by gold-plating stacks of micro-machined  silicon wafers.  This fabrication method is nearly identical to what was done for the previous generation of single frequency NIST horn coupled polarimeters \cite{Truce:britton} and is the most cost effective way to produce thermal coefficient of expansion matched ring loaded feeds for the millimeter band.

\begin{figure}
\begin{center}
\includegraphics[width=0.83\linewidth,keepaspectratio]{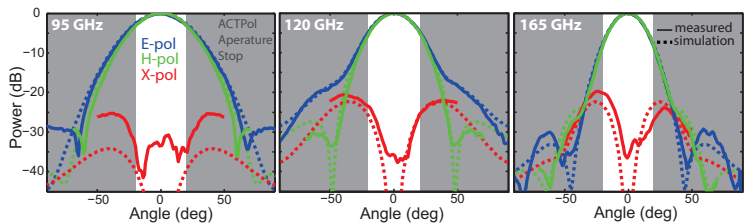}
\end{center}
\caption{Measurements and simulations of the beam pattern of the ring loaded corrugated feed horn.  These measurements of the E-, H-, and X-pol beams were made using a room temperature beam mapping system.  The angles at which the beams hit the ACTPol aperture stop are greyed out. (color figure online)}
\label{fig:hornmeaus}
\end{figure}

After passing through the horn, incoming radiation enters a detector holder that supports a planar detector chip in front of a waveguide back short and uses a waveguide choke to prevent leakage of fields from the waveguide.  The detector chip (See Figure \ref{fig:pixdesign}, right) uses a broad-band orthomode transducer\cite{multichroic:McMahon} (OMT) to couple the incoming light from the waveguide onto high impedance coplanar waveguide (CPW) lines. The OMT separates the incoming light according to linear polarization. The Y-polarized light is split onto the two vertically oriented OMT probes and propagate through identical electrical paths that have been highlighted in the figure.   Along each path, a broad-band CPW to micro-strip (MS) transition comprised of 7 alternating sections of CPW and MS is used to transition the radiation onto MS lines.   Next, diplexers comprised of two separate five pole resonant stub band-pass filters separate the radiation into 75-110 and 125-170 GHz pass-bands.  The signals from opposite probes within a single sub-band are then combined onto a single MS line using the difference output of a hybrid tee\cite{coupler:Knoechel}.   Signals appearing at the sum output of the hybrid are routed to a termination resistor and discarded.

These detectors operate over a 2.25:1 ratio bandwidth over which round waveguide is multimoded. However, the TE11 mode (which has desirable polarization properties) couples to opposite fins of the OMT with a $180^\circ$ phase shift while the higher order modes which couple efficiently to the OMT probes have a $0^\circ$ phase shift.   This fact allows the hybrid tee to isolate the TE11 signal at the difference port and reject the unwanted modes at the sum port.  This ensures single moded performance over our 2.25:1 bandwidth.  For testing purposes, the prototype pixel (shown in Figure \ref{fig:pixdesign}) included additional bolometers connected to the hybrid tee sum port.  The architecture described above offers excellent control over beam systematics of corrugated feeds, a frequency independent polarization axis defined by the orientation of the planar OMT, and a metal skyward aperture to minimize electrostatic buildup that is useful for future space applications.

\section{Multichroic Pixel Performance}
\label{sec:mcperf}

\begin{figure}
  \begin{minipage}[c]{0.55\textwidth}
    \includegraphics[width=0.88\textwidth]{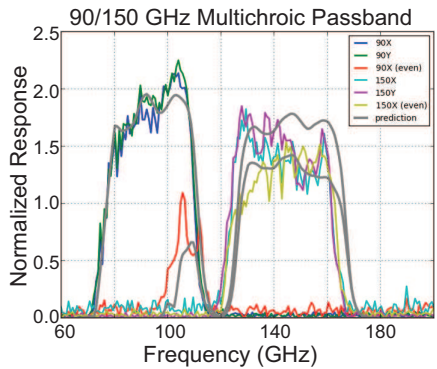}
  \end{minipage}\hfill
  \begin{minipage}[c]{0.45\textwidth}
    \caption{ \footnotesize FTS measurements and simulations of detector pass-bands.  For the 90 and 150 GHz channels, measurements of both linear polarizations as well as the terminated sum port were carried out.  The agreement between the sum port (even mode) measurements and simulations (gray lines) show that these detectors are properly rejecting unwanted waveguide modes. The agreement between the measured and simulated pass-band for the nominal X and Y polarized pixels is good, but has a 2\% shift in the band edges. (color figure online)} 
\label{fig:passband}
  \end{minipage}
\end{figure}

\begin{table}[b]
\caption{Losses in post-OMT microwave components}
\begin{center}
\begin{tabular}{|c|c|c|}
\hline
Item & 90 GHz loss & 150 GHz loss\\
\hline
CPW$\to$MS transition & 2.5\% & 4.1\%\\
 bandpass filter & 10.0 \% & 13.5\%\\
 hybrid tee & 3.5\% & 3.3\%\\
 microstrip line & 7.8\%& 13.1\%\\
\hline
total efficiency & 76\% & 66\% \\
\hline
\end{tabular}
\end{center}
\label{tab:eff}
\end{table}

Extensive measurements have been carried out on prototype pixels.  Figure \ref{fig:hornmeaus} shows measurements of the horn beam pattern made using a warm beam mapping facility. The horn was mounted on a three-axis (Az/El/phi) rotating stage and the measurements were carried out using a tunable microwave source and a diode detector. These measurements show better than 1\% agreement with theoretical predictions based on HFSS\cite{Ansoft:HFSS} simulations, and demonstrate round beams with low cross-polarization.  Calculations of the temperature to polarization leakage induced by these beams demonstrate that these pixels achieve temperature to polarization leakage subdominant to the CMB B-modes for $r = 0.01$.  Figure \ref{fig:passband} shows measurements and simulations of the detector pass-bands.   These results show excellent agreement with simulations for the 90 and 150 GHz passbands both for the  nominal polarization sensitive bolometers and the extra test bolometers connected to the sum ports of the hybrid tee.   This agreement verifies the function of the diplexer and confirms that unwanted modes are being correctly rejected by the hybrid tee.  Measurements of the optical efficiencies of these horn coupled detectors, carried out by placing 300 K and 77 K loads in front of a cryostat, show end-to-end optical efficiencies of 65\% and 50\% respectively for the 90 and the 150 GHz bands.  We estimate that the optical efficiency of the detectors are roughly 10\% higher after accounting for the losses in the non-AR-coated windows and filter stack in the measurement set-up.  These efficiencies are consistent with the predicted losses for the various detector components that are listed in Table \ref{tab:eff} based on $\tan \delta  = 0.005$ for the SiO$_{2}$ and  $\tan\delta  = 0.003$ for the SiN$_{x}$ \cite{SiNx:Cataldo} dielectric layers.  The agreement between simulations and measurements of the beam, pass-band, and optical efficiency demonstrate the function of the prototype detectors.  We are now working on incorporating this pixel design into a detector array for ACTPol.

\section{Array Optimization and Implementation on ACTPol}
\label{sec:arropt}

\begin{figure}[b]
  \begin{minipage}[c]{0.5\textwidth}
    \includegraphics[width=0.92\textwidth]{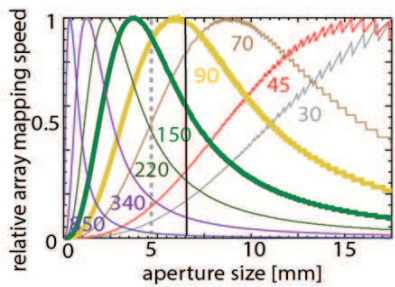}
  \end{minipage}\hfill
  \begin{minipage}[c]{0.5\textwidth}
    \caption{
        A simulation of the mapping speed as a function of horn aperture for a six inch detector array with 1.4:1 ratio bandwidth detectors with the band centers as labeled.   For a multichroic array with two wide bands it is possible to achieve sensitivity equivalent to 1.7 single frequency arrays.  Adding additional frequency bands leads to diminishing returns in sensitivity as these are sub-optimally coupled to the telescope. (color figure online)} 
\label{fig:mapspeed}
  \end{minipage}
\end{figure}

Maximizing the sensitivity of a fixed diameter detector array for a given optical system requires choosing the optimal horn aperture size.  If the aperture is too small, diffraction decreases the optical coupling to the telescope reducing sensitivity.  If the aperture is too large, then few detectors fit into the detector array and sensitivity is also reduced.  This tradeoff is evaluated quantitatively in Figure \ref{fig:mapspeed} which presents the mapping speed as a function of feed aperture for a six inch diameter detector array with the ACTPol optics.  This calculation takes into account a detailed optical model including transmission efficiency and emission from the stop, filters, optics, and atmosphere.  The calculation is performed for a variety of CMB bands each with 1.4:1 ratio bandwidth.  The peak of a single curve corresponds to the optimal horn aperture for a single frequency array.  For a multichroic array one must choose a horn aperture that maximizes the sensitivity of two or more channels simultaneously.  For the case of two neighboring bands (e.g. 90 and 150) it is possible to achieve 90\% of sensitivity of an optimized 90 GHz and an optimized 150 Ghz single frequency array with one multichroic array.   

The original plan for ACTPol was to field two 150 GHz arrays and a third 90 GHz array using the readout system from ACT that had 1024 channels per array.  Thus for the ACTPol multichroic array optimization we were restricted by the number of available channels which limited us to 256 feed horns with 4 detectors per pixel.  This drove us to a larger aperture (7 mm) than the true optimum for a multichroic array.  This leads to nearly optimal mapping speed at 90 GHz, but reduced sensitivity at 150 GHz.  Repeating the calculations shown in Figure \ref{fig:mapspeed} including the detector losses described in \S \ref{sec:mcperf} shows that this array will achieve 90\% the mapping speed of a 90 GHz array and 60\% the mapping speed of a 150 GHz single frequency array.  This comparison assumes lower losses for the single frequency arrays owing to the shorter MS lengths and use of free space filters instead of the stub filters.  With this array we gain a factor of 1.35 in sensitivity, which motivates the decision to switch from the baseline plan for the 90 GHz only array.  Moreover, this only results in a 4\% inflation of the 90 GHz map noise measured in ${\mu K}$.  In future applications of this technology we plan to increase the number of readout channels and optimize the optical design to optimally field multichroic arrays.  With this optimization a single multichroic array would be equivalent to 1.7 times a single frequency detector array in terms of sensitivity while also boosting spectral coverage.

\begin{figure}
  \begin{minipage}[c]{0.32\textwidth}
    \includegraphics[width=2.2\textwidth]{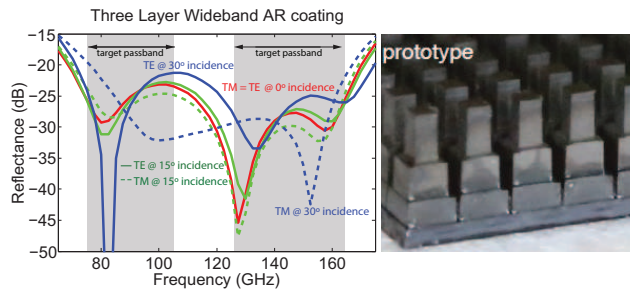}
  \end{minipage}\hfill
  \begin{minipage}[c]{0.25\textwidth}
    \caption{
     {\it Left:} A simulation of a wide bandwidth three layer metamatieral antireflection coating for silicon.  {\it Right:} A silicon prototype of this coating. (color figure online)} 
\label{fig:ARC}
  \end{minipage}
\end{figure}

Implementing multichroic arrays requires optical designs with the required wide bandwidth.   The limiting element for ACTPol optics is the antireflection (AR) coating for the three silicon lenses that make up the cryogenic re-imaging optics.  We have developed a new approach to AR coating cryogenic silicon lenses by cutting sub-wavelength features into the surfaces using silicon dicing saw blades \cite{lenses:datta}.  Figure \ref{fig:ARC} shows the simulated performance and a photograph of a mechanical prototype.  This coating will meet the requirements for fielding this multichroic array on ACTPol. A similar approach, consisting of conical holes drilled into UHMWPE, will be used for AR coating the receiver dewar window.

\section{Conclusions and Future Work}
\label{sec:conc}
We have presented the design of our horn coupled multichroic polarimeters and have shown data that verifies the performance of prototypes.  We are working on the development of a multichroic array with 255 feed horns and 1020 bolometers that will be deployed on ACTPol in early 2014.  This array is projected to achieve the sensitivity of 1.35 single frequency arrays.   Future arrays with fully optimized instruments can achieve performance equivalent to  1.7 single frequency arrays.  We have made prototypes of this design for higher frequency channels (180 to 360 GHz) which will soon be tested.  We are also developing a new design based on quad-ridge waveguide.  This new design will enable wider bandwidth (3.5:1 ratio bandwidth appears feasible) while substantially shortening the lengths of wires, eliminating the need for a hybrid tee and thereby reducing loss especially at the high frequency end of the pixel design.  

\begin{acknowledgements}
This work was supported by NASA through award NNX13AE56G and the NASA Space Technology Research Fellowship grant  NNX12AM32H and by the U.S. National Science Foundation through awards AST-0965625 and PHY-1214379.
\end{acknowledgements}


\begin{thebibliography}{99}

\bibitem{CMBPol:Bauman}
\small D. Baumann et al {\it AIP Conference Proc.} \textbf{1141}, 10-120, (2009)

\bibitem{CMBPol:Smith}
K. Smith et al {\it AIP Conference Proc.} \textbf{1141}, 121-178, (2009)

\bibitem{ACTPol:Niemack}
M. Niemack et al {\it Proc. of the SPIE.} \textbf{7741}, 21, (2010)

\bibitem{Ringloadedguide}
Y. Takeichi, T. Hashimoto and F. Takeda {\it IEEE Trans. Microwave Theory Tech.} \textbf{MTT-19}, 947-950, (1971)

\bibitem{Truce:britton}
J. Britton et al {\it Proc. of the SPIE.} \textbf{7741}, 11, (2010)

\bibitem{multichroic:McMahon}
J. McMahon et al {\it Journal of Low Temperature Physics} \textbf{167}, 5-6, (2012)

\bibitem{coupler:Knoechel}
R. Knoechel and B. Mayer IEEE \textbf{MTT-S} {\it International Microwave Symposium Digest} p. 471, (1990)

\bibitem{Ansoft:HFSS}
Ansoft {\it High Frequency Structure Simulator (HFSS)} software package

\bibitem{SiNx:Cataldo}
G. Cataldo et al {\it Optics Letters} Vol. \textbf{37}, No. 20, pp. 4200-4202, (2012)

\bibitem{lenses:datta}
R. Datta et al {\it Applied Optics} Vol. \textbf{52}, Issue 36, pp. 8747-8758 (2013)


\end{thebibliography}
\end{document}